\documentclass[10pt]{article}

\title{On the maximum drawdown during speculative bubbles}
\author{Giulia Rotundo$^{1,2}$\footnote{Corresponding author. $E-mail$ $addresses$:
giulia.rotundo@uniroma1.it (G. Rotundo), mauro.navarra@uniroma1.it
(M. Navarra)}, Mauro Navarra$^2$
\\  \normalsize $^1$Faculty of Economics,
University of Tuscia, Viterbo, Italy \\ \normalsize $^2$ Faculty of
Economics, \normalsize University of Rome  ``La Sapienza"}

\usepackage{graphicx}
\usepackage{rotating}

\begin{document}
\maketitle
\begin{abstract}A taxonomy of large financial crashes proposed in the literature
locates the burst of speculative bubbles due to endogenous causes
in the framework of extreme stock market crashes, defined as falls
of market prices that are outlier with respect to the bulk of
drawdown price movement distribution. This paper goes on deeper in
the analysis providing a further characterization of the rising
part of such selected bubbles through the examination of drawdown
and maximum drawdown movement of indices prices. The analysis of
drawdown duration is also performed and it is the core of the risk
measure estimated here.
\end{abstract}

\noindent{\em Key words} Risk measure, drawdown, speculative
bubbles.
\newline {\it PACS} { 89.65.Gh Economics, business, and financial markets, \newline 89.90.+n
Other areas of general interest to physicist }

\section{\label{sec:introduction}Introduction}
  An insight in the long term behavior of
portfolios is a  delicate task in long term investment strategies.
The need to consider extreme financial market events encompasses
the investigation of large financial crashes, that were already
classified as outliers with respect to the bulk of market drops,
and often associated to the burst of speculative bubbles due to
endogenous causes \cite{js}. \newline This paper
aims at 
extracting risk
features that characterize the rising part of such speculative
bubbles.

\begin{figure} [htbp]
\centering
\includegraphics[totalheight=14cm,width=12cm]{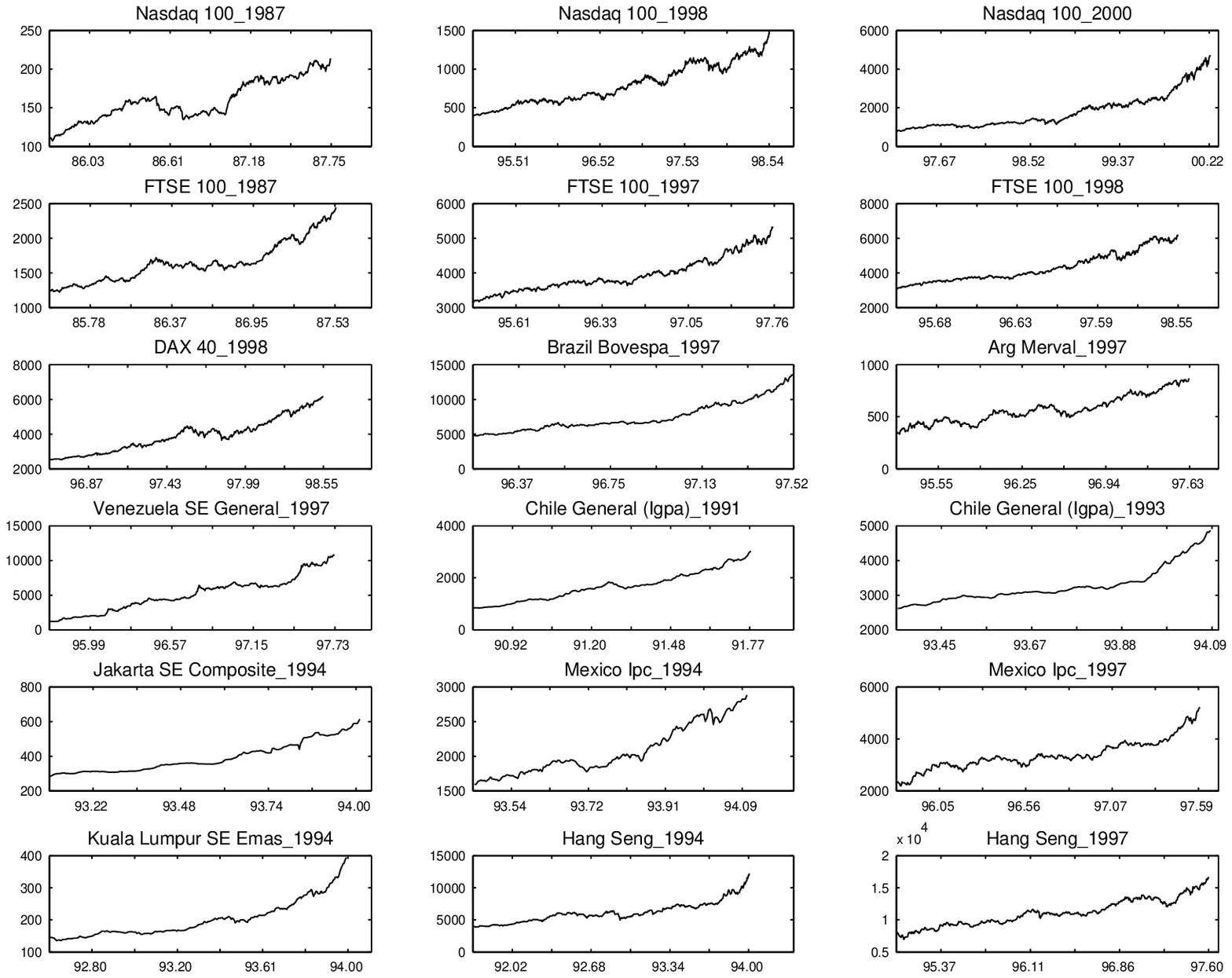}
\caption{Rising part of speculative bubbles selected. Large
financial crashes were chosen in accord to \cite{js}. The complete
list is given in Tab. \ref{tab:crashes}. Time is reported on $x$
axis, in accord with the notation used by \cite{js}. $y$ axis
reports the value of each index.} \label{fig:dataset1}
\end{figure}

 Our data selection relies
on the huge data analyses  worked out by Johansen and Sornette
\cite{js,sorL}. In their several papers, they develop and support
through empirical evidence  the theory describing speculative
bubbles due to endogenous causes like as systems close to some
rupture point. In particular large market indices drops ending
speculative bubbles due to endogenous causes have been characterized
through the occurrence of log-periodic-power laws (LPPL),
interpreted as a signature of underlying cooperative phenomena among
market agents (Fig. \ref{fig:dataset1}, Tab. \ref{tab:crashes}).
These results were related to analyses of drawdown movements of
market indices prices (DD), thus providing systematic taxonomy of
crashes \cite{js,jocar,js3}.
Our analysis starts from this point and goes on inquiring drawdown
market prices movements duration and size. The analysis of duration
of drawdown movement is relevant in itself, considering the key role
of time in catastrophic events like as large financial crashes are,
that show a drop of prices within short time intervals
\cite{js,sorL,joh}. The analysis of drawdown size requires as a
preliminary step the selection among the several different
definitions of drawdown that were proposed in literature. Results
depend strictly on the measure chosen for drawdown, and on the
actual data set \cite{js,joh,jsoutliers,mendes}. A first analysis
classified large stock markets indices crashes as outliers whether
considering the stretched exponential distribution of market drops
\cite{jsoutliers}. A further analysis for noise filtered drawdowns
allowed to calibrate the parameter in order to avoid outliers
\cite{joh}. Other studies using different definitions of drawdowns
and considering different data sets proposed  for returns drawdown
probability density fitting stable models \cite{miao}, the
generalized Pareto distribution (GPD) \cite{mendes,miao}, and the
modified GPD. The worst scenario of maximum drawdown movements (MDD)
has also been examined. Results depend also on the time window on
which the maximum is estimated.
For example in \cite{mendes} the time window selection corresponds
approximately to half an year, the entire data sets cover more than
20 years, and the results are used for probability distribution
calibration. In \cite{miao} moving windows of 63 data each are used
on datasets including at least 15 years. In both cases the long time
series of market indices data that were used include lower and
bigger financial crashes, the rise and the deflate of bubbles, till
to the return to fundamentals, and the several changes of regimes
included in societal and trading dynamics spanning the last 15-20
years. Therefore, the maximum drawdown measure calculated on moving
windows gives samples of a process obeying several different
dynamics. Here we aim at using a different approach for drawdown
modeling. Instead of splitting the entire time series that are
available, we extrapolate the probability of drawdown and maximum
drawdown during the rising part of the speculative bubble, looking
for the extraction of common features.\newline We compare the
results obtained on the DD through the estimate of risk measures.
 Risk
exposure in financial markets has been described through several
measures (VaR, CVaR, etc.) based on statistical properties of data
that do not consider the order of data sequences. In fact empirical
mean, variance, as well as higher order moments of probability
distributions  are invariant under data shuffle. But long downward
trends containing long lasting sequences of consecutive drawdown
price movements could suggest investors to withdraw from the market,
and they can quite force small investors to such a choice
\cite{cuz}. Therefore, measures of risk based both on the duration
of consecutive market drops and on the maximum drawdown can play a
relevant role in driving investment strategies.

  We stress again that we use here a methodological approach
 deeply different form other comparative studies on risk measures,
 because we aim at extracting features of periods
(the rising part of speculative bubbles due to endogenous causes),
that are supposed to be driven by the same kind of dynamics, and
we proceed comparing them with the entire data set available,
whilst other studies that compare risk measures
 on long time windows do not take into account the evolution of
 dynamics driving stock markets across decades \cite{miao}.

\section{Large stock market indices crashes}
\label{sec:3} \hspace{6pt} A series of papers about speculative
bubbles proposes an explanation of large financial crashes due to
causes endogenous to the market. Similarities with critical
phenomena like earthquakes and the sound emission in materials close
to the rupture point led the research to the detection of
cooperative underlying phenomena evidenced through discrete scale
invariance in financial data (FX, Gold, stock market indices) close
to large crashes that can thus be well described as critical points
\cite{sorL}. In the case of market indices the logarithm of index
values close to the crash time is described by a characteristic
LPPL:
\begin{equation} \label{Iorder}
F(t)\simeq A+B(t_c-t)^{m} + C (t_c-t)^{m} cos( \omega ln(t_c-t)-
\phi) \,\,\,\,\,\,  t<t_c  \,\,
\end{equation} where $t_c$ is the crash time, and $A$, $B$, $C$, $m$ , $\omega$, $\phi$, and $t_c$
are parameters to be estimated via numerical optimization.
  A further taxonomy  that distinguishes the causes that
generate crashes can be found in \cite{js} and confirms the
detection of LPPL as a hallmark of speculative bubbles due to
causes endogenous to the market \cite{js,abmv2,abmv}
cross-validated with the amplitude of the crash ending the bubble
as an outlier from probability distribution of the bulk of price
variations
 \cite{js3,jsoutliers}.
%
 Stock market indices datasets reported in Tab. \ref{tab:crashes} were selected lying on the results of \cite{js} in which the
selection of an outlier market drop is used in itself as a
definition of crash and then it is  cross-validated with the
occurrence of LPPL. The rising period of the bubble in which the
LPPLs occur is selected starting from a lowest point before the rise
of the bubble and ending at the maximum value of the index
\cite{sorL}.

\section{Drawdown duration distribution}

 The time plays a key role in crash definition
\cite{js,sorL} and in the reaction of investors to the change of
market condition, but most attention up to now was paid to the MDD
than to the time in which it occurs,  although the time measure was
proposed as a first step towards the definition of some ``drawdown
velocity" \cite{js}. Therefore, for each selected data set we
exploit the duration of drawdowns, defined in the most simple way as
the number of time steps from a local peak to the next local
minimum, corresponding to the so called ``pure" drawdowns of
\cite{js,sorL,joh}.

\begin{figure}[hbp]
\centering
\includegraphics[totalheight=6cm,keepaspectratio,bb=0 0 384 288, ]{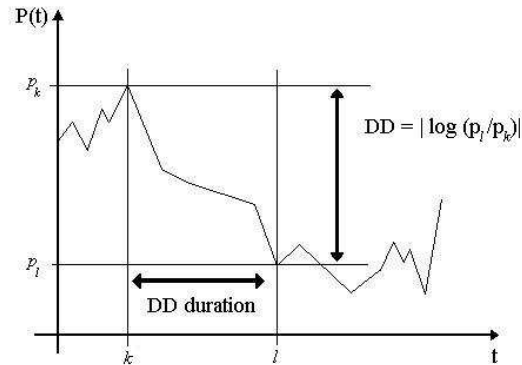}
\caption{Pure DD definition in accord to (\ref{DDdefinition})}
\label{fig:DD definition}
\end{figure}

\begin{figure} [htbp]
\centering
\includegraphics[totalheight=14cm,width=12cm]{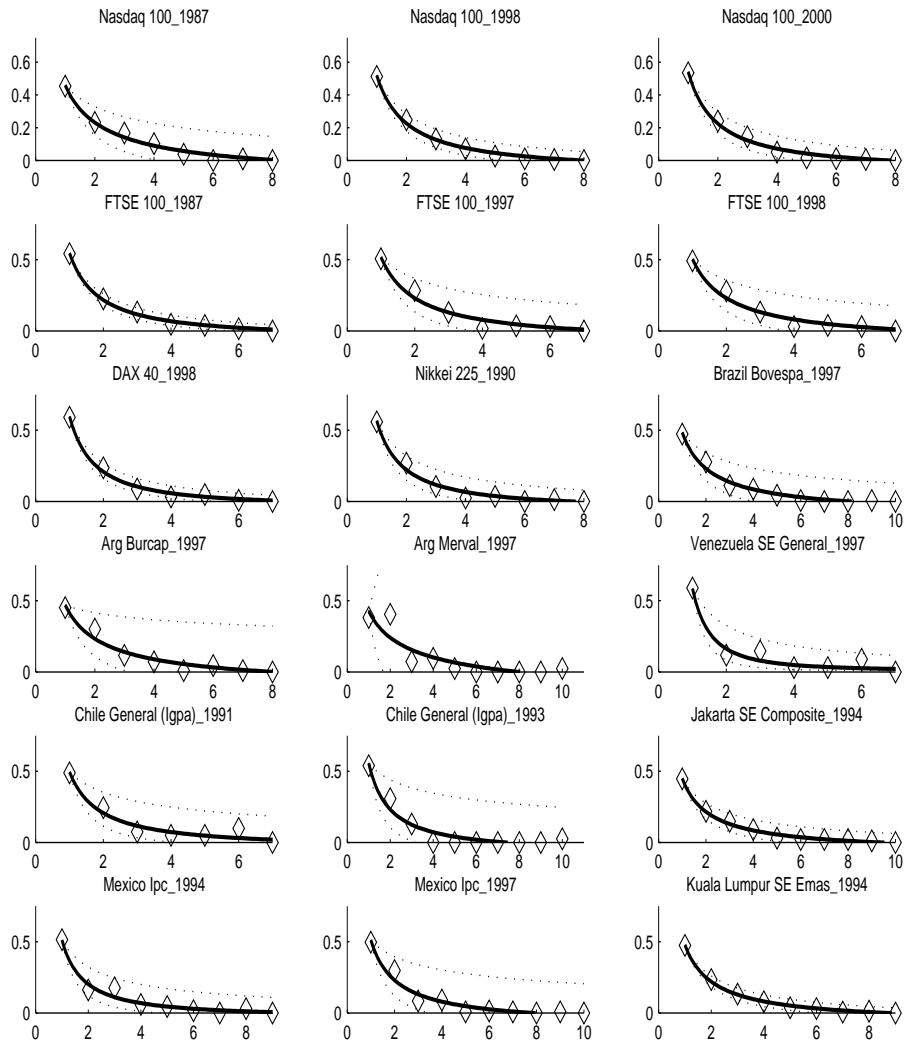}
\caption{Histogram of the duration - rising parts of the speculative
bubbles.  Usual histogram bars were substituted by diamonds because
of graphic clarity. $x$-axes report the DD duration in term of days;
$y$-axes report their frequencies. On each example the solid line is
the power law regression curve; the dotted line is drawn using the
bounds of the 95\% confidence intervals. The decay exponents $\alpha
\in [0.46, 1.7]$ where 0.46 is the minimum value and 1.7 is the
maximum value over all the reported samples; the mean value of
$\alpha$ is approximately at $1$.}
\label{fig:IstoDuration7X3_Dataset1Rel}
\end{figure}

 A first empirical result for drawdown duration is shown in \cite{mendes}, where the negative binomial or the Poisson distribution is proposed  for modeling
drawdown duration, and the Gamma or Pareto distribution for
modeling the maximum drawdown. \newline Let $\{p_t\}_{t=1,T}$   be
a time series of a stock market index daily closure price, and let
$r_t=log(p_t/p_{t-1})$ be the usual log-returns. 
  The
following definitions are in accord with the drawdown definition
used in \cite{mendes}, apart from a normalization factor that is not
relevant for our analysis (Fig. \ref{fig:DD definition}). \newline
 {\bf Definition 1} Let $p_k$
be a local maximum, and  $p_l$ be the next local minimum:
\begin{equation} \label{seq} p_k>p_{k+1}> \cdots >p_l, \, \mbox{with} \,\, l-k
\ge 1, k \ge 1, \mbox{and} \, p_k>p_{k-1},\, p_{l+1}>p_{l}.
\end{equation}  A pure DD (size) is defined as
the partial sum of daily returns
\begin{equation} \label{DDdefinition}DD_{k,l}=\mid r_{k+1}+r_{k+2}+ \cdots
r_{l} \mid= \mid \sum_{j=1}^{l-k} r_{k+j}\mid = \mid log
\frac{p_l}{p_k}\mid  \end{equation} {\bf Definition 2} The drawdown
duration is the time length  $l-k$ of the sequence of negative
returns defined through (\ref{seq}).\newline DD duration was
estimated on each selected time series. The power law was fitted to
DD duration histograms. On the sets of the rising parts of selected
bubbles shown in Fig. \ref{fig:dataset1} the power law exponent
ranges from 0.46 to 1.7, having a mean value $\alpha \simeq 1$. On
the entire data sets $\alpha$ ranges from 0.67 to 1.10, with
$\alpha$ mean value $\alpha \simeq 0.94$, accordingly to the
presence of longer and deeper crashes in the enlarged time domain.
The Poisson distribution gives a poor fit of DD duration histogram
on the lowest duration values, and no useful result comes from the
negative binomial fit. Therefore, we are going to use the empirical
estimates for the risk measure computation (Fig.
\ref{fig:IstoDuration7X3_Dataset1Rel}).

Studies on speculative bubbles have attempted to extract common
features beyond LPPL. A first attempt \cite{mendes} to relate GARCH
parameters with the prediction of the volatility connected to the
drawdown movement is made clearer in \cite{dsz}: the presence of a
bubble drives the high volatility in the GARCH model, but the
reversal implication does not hold. The power law exponents of
drawdown duration decay seem to evidence some general feature of
drawdown duration interesting in itself.
\newline
The grouping of the exponents in a range supports the hypothesis of
close risk profiles across international markets, form the DD
duration point of view. However, a further correlation analysis
comparing the $\alpha$ estimated on each series and the LPPL
paramenters $\omega$ and $z$ does not show significative empirical
correlation, like at it was stated for GARCH parameters.


\section{Relationships with other drawdown definitions}
The delicate task of defining what is a crash, a large drop or a
significant change of regime is far from being well assessed
\cite{dsz}. Also the closely connected definitions of drawdown
market movements analyzed in the literature are not homogeneous
\cite{js,cuz,mia}. Therefore we are going to examine the
relationship between the so called {\em pure drawdown} (DD), {\em
$\epsilon$-drawdown} ($\epsilon$DD) \cite{js,mendes}, that are used
in the theory of speculative bubbles, and the maximum drawdown
($\bar{D}$) used in \cite{cuz,mia}, that provides a risk measure
interestingly related to the Calmar and Sharpe ratio.
\newline {\bf Definition 3} The maximum drawdown $M$ is the longest partial sum
of daily returns
\begin{equation} M=max\{ DD_{k,l} \mid \mbox{conditions (\ref{seq}) hold} \}
\end{equation}
The definition of $\epsilon$DD relaxes the condition (\ref{seq})
allowing for small rises of values during the decreasing sequence.
This description can be formalized through the following set of
conditions
\begin{equation} \label{seq2} \begin{array}{l}
k=arg ( max_s (s-l)), \, s.t. \,\, p_k>p_{k+1}
> p_{k+2} -\epsilon> \cdots
>p_l -\epsilon , \,    l-k \ge 1, k \ge 1,\,\,   \\ p_k>p_{k-1}, p_k>p_{i}, i=1,...,l, p_{l+1} > p_l + \epsilon, \epsilon \ge 0.\end{array} \end{equation}
and therefore to the following
\newline
 {\bf Definition 4} The  $\epsilon$DD ending at time $l$ is given by
\begin{equation} \epsilon DD_l=\{DD_l \mid  \mbox{conditions (\ref{seq2}) hold} \}. \end{equation}
 This definition states that price growth below a certain magnitude $\epsilon$ is ignored, and it serves to
filter noise. If $\epsilon=0$ the definition corresponds to  ``pure"
drawdown;  filter size was chosen as a function of the empirical
volatility $\sigma$ ($\epsilon=0, \sigma/4, \sigma/2$) \cite{js}.
This definition includes in the particular case $\epsilon=0$ the
definition of pure drawdown ending at time $t$ ($DD_t$).
\newline
 {\bf Definition 5}  The duration of $\epsilon DD_l$
 is given by
\begin{equation} \label{seq2duration}
max_s \{(s-l) \mid  \mbox{conditions (\ref{seq2}) hold} \}
\end{equation}
$\bar{D}$ will be the worst loss, i.e. max of the above.
 \newline The definition used in \cite{mia} starts from a
stochastic processes approach:
\newline
 {\bf Definition 6} Let $\{X(t)\}$ a stochastic process. $\bar{D}$ is given by
\begin{equation} \label{mdd}
\bar{D}=sup_{t \in [0,T]} ( sup_{s \in [0,t]} X(s) - X(t)).
\end{equation}
This measure gives the range from the maximum to the minimum anytime
the maximum precedes the minimum.  $\bar{D}$ was calculated using
$X(t)=log(p_t)$, so to deal with returns.
\newline Although the definition of $\epsilon$DD was aimed only at
filtering the noise the following remark holds:
\newline
 {\bf Remark 1} Let $X(s)=log(p_s)$. For each $t$ ending time of an $\epsilon$DD$_t$:
\[ \epsilon DD_t <\bar{D}\]
 {\bf Remark 2}
$\bar{D}$  is  an $\epsilon$DD$_t$ for $X(t)=log(p_t)$, $\epsilon
> max_t \mid p_t-p_{t-1} \mid$.
\newline
 Henceforth risk measures based on DD and $\bar{D}$
will provide bounds to $\epsilon$DD. The following chain of
inequalities is a direct consequence of the weakening of conditions
in the definitions.
\newline
 {\bf Lemma 1} In accord with the definitions reported above:
\begin{equation}\begin{array}{c} DD_t < \epsilon DD_t   < \bar{D} \\
M   < \bar{D} \end{array}
\end{equation}

The size of drawdowns described by $\bar{D}$ is higher of $M$, due
to the weakening of the descent conditions, and provides a worse
scenario. \newline The same remark allow to conclude that the same
inequality holds on the duration of DD$_t$ and $\epsilon$DD$_t$.

\section{A measure of risk based on drawdown movements size and duration}

\begin{figure}[htbp]
\centering
\includegraphics[angle=90,height=16cm,width=12cm]{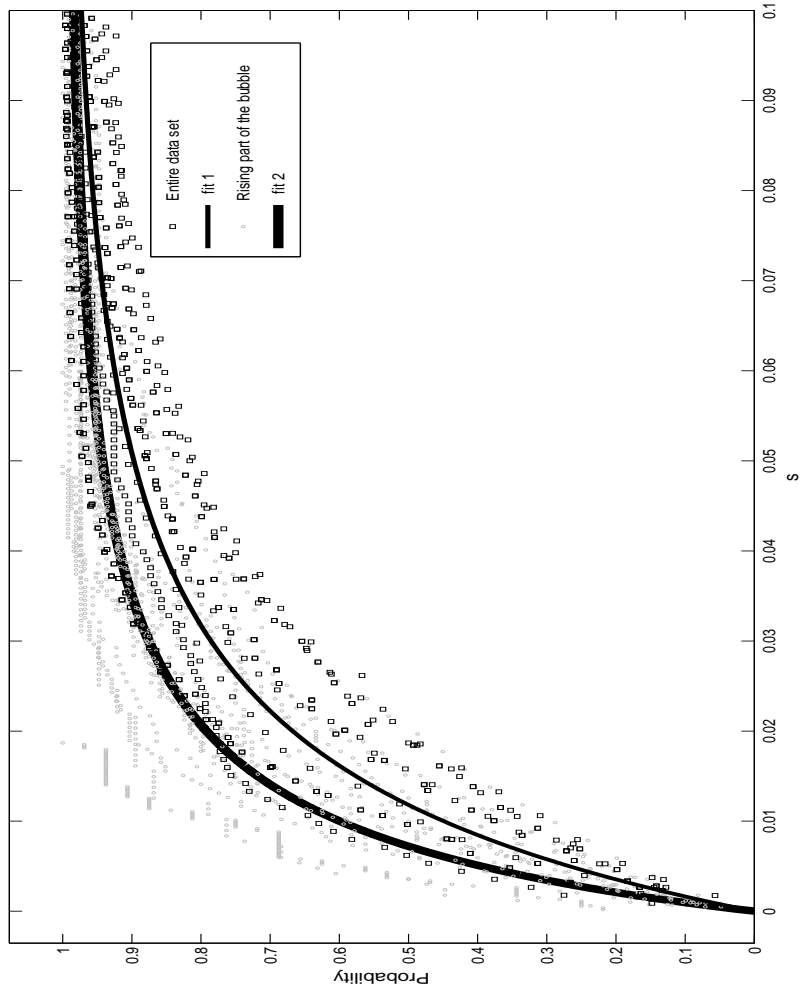}
\caption{Best fit of empirical estimates of (\ref{mendesDD}) through
(\ref{gpd}). $x$-axis reports the values of $s$; $y$-axis the
cumulative probability. The curve labeled  ``fit1" is the regression
curve obtained by fitting (\ref{gpd}) on the empirical estimates of
(\ref{mendesDD}) on the set of entire time series. Parameters values
are $\beta= 0.0153 (0.0148, 0.0158)$, $\xi =0.3012 (0.2698,
0.3326)$. The curve labeled  ``fit2" is the regression curve
obtained by fitting (\ref{gpd}) on the empirical estimates of
(\ref{mendesDD}) on the set of rising part of bubbles. Parameters
values are $\beta= 0.0089 (0.0085, 0.0093)$, $\xi = 0.4273 (0.3783,
0.4763)$. Values reported in parentheses are 95\% confidence
interval.} \label{fig:GPD}
\end{figure}

\begin{figure} [htbp]
\centering
\includegraphics[height=16cm,width=12cm]{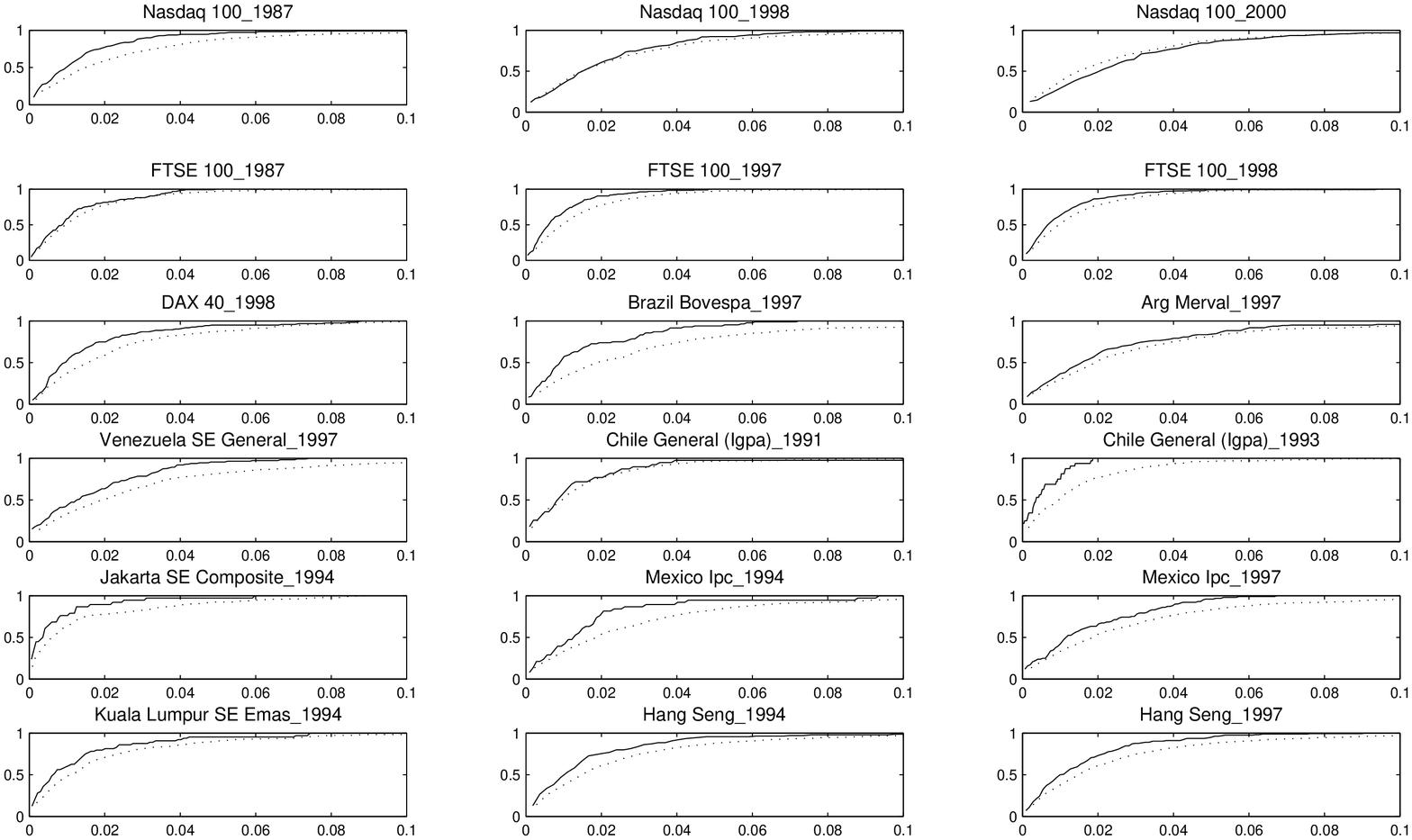}
\caption{Empirical estimates of (\ref{mendesDD}). $x$-axes reports
the value of $s$; $y$-axes the cumulative probability. Each figure
reports (\ref{mendesDD}) estimated on the rising part of a selected
bubble (solid lines) and compared with (\ref{mendesDD}) estimated on
the entire time series (dashed lines).}
\label{fig:Distribuzionediprobcumulate6X3}
\end{figure}

Stock market indices actually are a particular portfolio, basing
on a
 weighted mean of selected stock prices, and to buy/sell
 stock market indices has the
 meaning to buy/sell a previous selected financial product replica of
 the index (Exchange Traded Funds, certificates).
Portfolio risk measures considering DD and MDD should be used in a
complementary way with respect to the traditional ones (VaR, ES)
at least in the case of Stable Paretian distribution \cite{miao}.
 This section aims at comparing
the behavior of a risk measure based on drawdown movement size  on
the rising part of speculative bubbles and on the entire time
series.
 Results of \cite{js} that
classify as outliers with respect to the stretched exponential
distribution the large financial crahes rising from the burst of
speculative bubbles due to endogenous causes were extended
considering coarse-grained drawdowns ($\epsilon$-drawdowns)
\cite{joh}, and later the GPD distribution for negative tails was
tested \cite{mendes,miao,brandi,mps}. These last results go beyond
the mere distribution hypothesis testing, relying on literature on
extreme events, that proposes the GPD, as a universal description of
the tail of distributions of Peaks-Over-Thresholds. The same
approach can be used for the MDD estimates. Although the duration
plays a key role in drawdowns little or no analysis has been
reported on the literature on drawdown duration \cite{mendes}.
Therefore, we perform analyses on the joint probability of both
drawdown size and drawdown duration. The following measure of
drawdowns
\begin{equation} \label{mendesDD}
Pr\{DD<s\}= \sum_{d=1}^\infty Pr\{DD_{k,l} <s \ \mid l-k=d \}
Pr\{l-k=d\}
\end{equation}
is estimated on the rising part of each selected bubble and on the
entire time series of the corresponding stock market index. This
approach differs from the one used in \cite{mendes,brandi}, that
relies on probability obtained as a best fit of empirical drawdown
size distribution, but that is based on independence hypothesis for
drawdown duration modeling.
\newline
The weight of the biggest crashes in the entire time series
evidences in lowering curves corresponding to entire time series.
\newline
Following the approach of \cite{mendes} we fit the GPD
\begin{equation} \label{gpd} G_{\xi,\beta} (x)= 1-(1+\xi*x/\beta)^{\frac{-1}{ \xi}} \,\, , \,\,\, x \in D(\xi, \beta) \end{equation}
 on the set
of (\ref{mendesDD}) estimated  on the rising
 part and on the set of  entire time series (Figs \ref{fig:GPD}, \ref{fig:Distribuzionediprobcumulate6X3}).
This result provides the   mean behavior  of drawdown to which refer
in case of speculative bubbles, and evidences the difference with
respect to the common behavior of the entire time series. On the
rising part of the bubble  $\beta=  0.0089  (0.0085, 0.0093)$, $\xi
= 0.4273 (0.3783, 0.4763)$; on the entire time series $\beta= 0.0153
(0.0148, 0.0158)$,   $\xi =0.3012 (0.2698, 0.3326)$. Values reported
in parenthesis are 95$\%$ confidence interval. The ranges of $\xi$
and $\beta$ parameters are not overlapping.
 The
change of parameters value gives also another measure of large drops
considered in the entire time series if compared with the drops
experienced during the bubble. For each selected bubble Fig.
\ref{fig:Distribuzionediprobcumulate6X3} shows the cumulative
function (\ref{mendesDD}) of both the rising part of the bubble and
the entire index. The closest the functions, the closest the
drawdown structures are, and the lowest the impact of the outlier,
that implies also that  minor crashes in the rising part of the
bubble have a relevant weight. As an example, drops during the
rising part of the speculative bubble of Nasdaq collapsed in April
2000 had an intensity close to the Nasdaq crashes of the previous
years, leading to highly overlapping curves.\newline Referring to
Fig. \ref{fig:GPD},  ``fit 2" is the regression on the sets of
(\ref{mendesDD}) estimated on the rising part of speculative
bubbles.  ``fit 2" separates DD behavior on the entire data set from
1/3 of (\ref{mendesDD}) estimated on the rising part of speculative
bubbles, corresponding to series exhibiting smaller and/or shorter
DD.
\newline Risk profiles of rising part/entire time series are not
completely separated, but the result allow to use lower risk
profiles in the rising part of speculative bubbles whether carefully
encapsulated in a predictive scheme of burst of bubbles like
\cite{dsz}.

\section{Maximum of drawdown movements}
\begin{figure} [htbp]
\centering
\includegraphics[height=6cm,width=12cm,keepaspectratio=true]{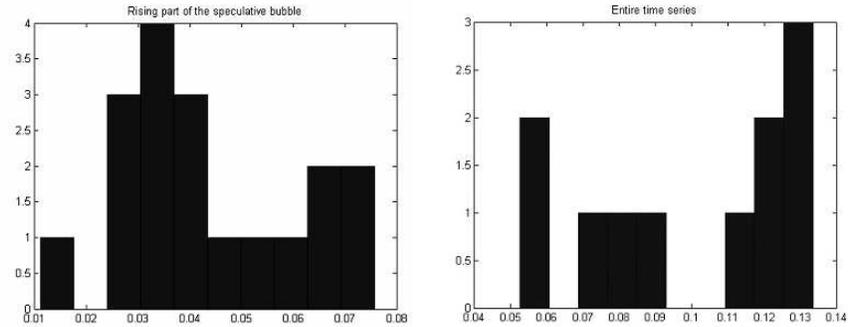}
\caption{Histogram of maxima of pure drawdown (\ref{mendesMaxDD}).
$x$-axes reports values of $E[M]$; $y$-axes reports the counting of
them. On the set of rising part of speculative bubbles the range is
$(0.01,0.08)$; on the set of entire time series the range is
$(0.05,0.13)$. } \label{fig:maxDDcond}
\end{figure}

\begin{figure} [htbp]
\centering
\includegraphics[height=6cm,width=12cm,keepaspectratio=true]{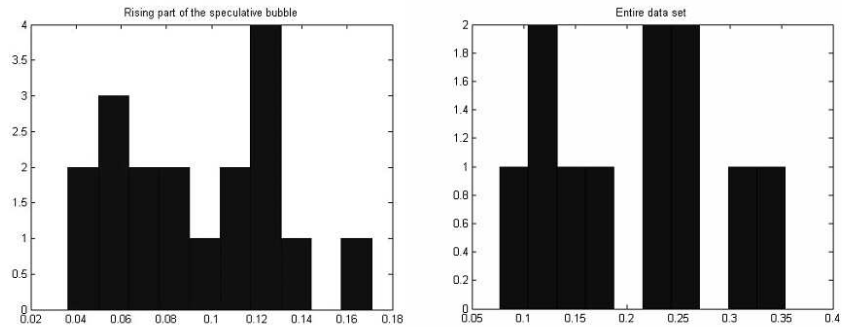}
\caption{Histogram of $\bar{D}$ (\ref{mdd}). $x$-axes reports values
of $\bar{D}$; $y$-axes reports the counting of them.On the set of
rising part of speculative bubbles the range is $(0.04,0.17) $ ; on
the set of entire time series the range is $(0.08,0.35)$. }
\label{fig:istoMDD}
\end{figure}

Other measures of drawdown behavior were considered, in the case of
the occurrence of the worst scenario. A discussion of coherence
properties of measures DD-based is carried on \cite{mendes,cuz}.
$Pr\{M<s\}$ was calculated in \cite{mendes} conditioning to the
duration, like in (\ref{mendesDD}). The probability distribution of
$M$ was extracted through a segmentation of 15-20 years long time
series. This procedure relies on the (undeclared) assumption that
data in time windows considered obey a homogeneous underlying
process, that of course is a very raw assumption. The need to split
windows is due to the fact that only one $M$ is available for each
time series, but several samplings are needed in order to build a
distribution. Here we follow a deeply different methodological
approach, aiming at extracting features that can be shared by the
rising part of speculative bubbles. Therefore we consider
$max(DD_{k,l})$ conditioned to the duration and we estimate and
compare
\begin{equation} \label{mendesMaxDD}
E[M]=  \sum_{d=1}^\infty max\{DD_{k,l}\mid l-k=d \} Pr\{l-k=d\}
\end{equation}
on each time series. Results are shown in Fig. \ref{fig:maxDDcond}.
The leftmost picture reports the results on the rising part of the
bubble, the rightmost on the entire time series. This estimate
compares the size of maximum sequences of drops inside the bubble
with the ones outside it, allowing to understand the impact of a
huge financial crash with respect to the smaller crashes internal to
the bubble. $ E[M] \in (0.01, 0.08)$ (rising part) and $E[M] \in
(0.05, 0.13)$ (entire series). Ranges are partially overlapping,
remarking that some crashes classified as huge in some indices had a
magnitude not so impressive for other indices.
%
\newline Referring to Fig. \ref{fig:maxDDcond}, the discriminant analysis
locates $E[M] \in (0.01, 0.05)$ for the rising part of speculative
bubbles and $E[M] \in (0.08, 0.13)$ for the entire time series.
Intermediate values (0.05, 0.08) report the situation for the
crashes of Nasdaq 100 in 1998 and 2000, DAX 40 in 1998, Argentina in
1997, Hong Kong in 1994 and 1997 (rising parts of the bubbles; on
Fig. \ref{fig:dataset1} labeled, respectively, Nasdaq 100\_1998,
Nasdaq 100\_2000, DAX 40\_1998, Arg Merval\_1997, Hang Seng\_1994
and Hang Seng\_1997), and for FTSE 100, Chile General (Igpa),
Jakarta SE Composite (entire time series).\newline Fig.
\ref{fig:istoMDD} shows $\bar{D} \in (0.04, 0.17)$ for the rising
part of speculative bubbles; $\bar{D} \in (0.08, 0.35)$ for the
entire time series.
\newline The weakening of the descent conditions can also be
addressed for the higher overlap in windows values. Anyway this
measure is worth of being considered because it gives the very worst
loss case in case of long time buy-and-hold strategies.

\section{Conclusions and further developments}

This paper focuses on risk measures based on drawdowns and has
introduced the empirical estimate of duration in the loss function
underlying a risk measure. The analysis of the duration of drawdown
is relevant for fund managers, that can bear high volatility
periods, but that risk to lose clients whether a large sequence of
drops happens. Empirical results are used to extract characteristics
common to the rising part of speculative bubbles and to evidence
similarities and differences with respect to the entire time series.
A comparison between risk measure DD-based shows the improvement in
the discriminant analysis obtained by the introduction of DD
duration. The approach used here is deeply far from the most common
usage to split long time series or to simulate them basing on the
raw hypothesis of a homogeneous process underlying long time series.
Recently an alert system \cite{dsz} was proposed in order to open
the way to practical usage of the LPPL bubble theory for the
forecast of the bubble ending time. We look forward for the
embedding the present analysis into the predictive scheme of
\cite{dsz} for the calibration of investment strategies during the
rising part of bubbles, before the expected crash time.


\section*{Acknowledgements}
GR thanks her colleague Annamaria d'Arcangelis for useful
discussion on stock markets.

\begin{table} [htbp]
\caption{List of crashes. Data sets of rising part of speculative
bubbles are chosen since the rise of the bubble to the expected
crash time \cite{sorL}} \label{tab:crashes} \centering \vskip 1 em
{\small
\begin{tabular}{llll}
\hline\noalign{\smallskip}   Index\,\,\,\, & Crash year & Rising
part & Entire dataset
\\ \noalign{\smallskip}\hline \hline
Nasdaq 100 & 1987 &    10/01/1985 - 10/02/1987 &  10/01/1987 -
12/31/2001
 \\ \noalign{\smallskip}\hline 
Nasdaq 100 & 1998 &   01/02/1995 - 07/20/1998  & 10/01/1987 -
12/31/2001
 \\ \noalign{\smallskip}\hline
Nasdaq 100 & 2000 & 04/01/1997 - 03/27/2000  &  10/01/1987 -
12/31/2001
 \\ \noalign{\smallskip}\hline
FTSE 100 & 1987 & 07/01/1985 - 07/16/1987  &  07/01/1985 -
12/31/2001
\\ \noalign{\smallskip}\hline
FTSE 100 & 1997 & 04/03/1995 - 10/03/1997  & 07/01/1985 -
12/31/2001
\\ \noalign{\smallskip}\hline
FTSE 100 & 1998 & 03/14/1995 - 07/20/1998  & 07/01/1985 -
12/31/2001
 \\ \noalign{\smallskip}\hline
DAX 40 &1998 & 08/06/1996 - 07/20/1998  & 08/06/1996 - 12/31/2001
 \\ \noalign{\smallskip}\hline
Brazil Bovespa & 1997 &03/06/1996 - 07/08/1997 & 03/06/1996 -
12/29/2000
\\ \noalign{\smallskip}\hline
Arg Merval& 1997 &03/15/1995 - 08/20/1997& 03/15/1995 - 12/29/2000
\\ \noalign{\smallskip}\hline
Venezuela SE Gen& 1997 &09/12/1995 - 09/23/1997 & 09/12/1995 -
12/29/2000
\\ \noalign{\smallskip}\hline
Chile General (Igpa) & 1991 & 10/10/1990 - 10/08/1991 & 10/10/1990
- 12/31/1996
\\ \noalign{\smallskip}\hline
Chile General (Igpa) & 1993 & 05/10/1993 - 02/04/1994& 10/10/1990
- 12/31/1996
\\ \noalign{\smallskip}\hline
Mexico Ipc& 1994 &06/15/1993 - 02/08/1994 & 06/15/1993 -
12/29/2000
\\ \noalign{\smallskip}\hline
Mexico Ipc&   1997 &10/18/1995 - 08/06/1997& 06/15/1993 -
12/29/2000
\\ \noalign{\smallskip}\hline
Kuala Lumpur SE Emas& 1994 &08/06/1992 - 01/04/1994 &08/06/1992 -
21/31/1996
\\ \noalign{\smallskip}\hline
Hang Seng& 1994 &09/12/1991 - 01/04/1994 &  09/12/1991 -
12/29/2000
\\ \noalign{\smallskip}\hline
Hang Seng&1997& 01/02/1995 - 08/07/1997 & 09/12/1991 - 12/29/2000
\\ \noalign{\smallskip}\hline
\end{tabular}
}
\end{table}

\end{document}